\documentclass[aip,jmp,reprint,amssymb,amsmath]{revtex4-1} 
\usepackage{natbib}
\usepackage{subfig}
\usepackage{graphicx}
\usepackage{bm}
\usepackage{amsmath}
\usepackage{float}
\usepackage{caption}
\usepackage{ragged2e}

\begin{document}


\title{Time-resolved double-resonance spectroscopy: Lifetime measurement of the $6\,^1\Sigma_g^+(7,31)$ electronic state of molecular sodium}
\author{Michael Saaranen}
\author{Dinesh Wagle}
\affiliation{Department of Physics, Miami University, Oxford, Ohio 45056, USA}
\author{Emma McLaughlin}
\author{Amelia Paladino}
\author{Seth Ashman}
\affiliation{Department of Engineering, Physics, and Systems, Providence College, Providence, Rhode Island}
\author{S. Bur\d{c}in Bayram}
\email{bayramsb@MiamiOH.edu}
\affiliation{Department of Physics, Miami University, Oxford, Ohio 45056, USA}

\date{\today}

\begin{abstract}
We report on the lifetime measurement of the $6\,^1\Sigma_g^+ (7,31)$ state of Na$_2$ molecules, produced in a heat-pipe oven, using a time-resolved spectroscopic technique. The $6\,^1\Sigma_g^+ (7,31)$ level was populated by two-step two-color double resonance excitation via the intermediate $A\,^1\Sigma_u^+ (8,30)$ state. The excitation scheme was done using two synchronized pulsed dye lasers pumped by a Nd:YAG laser operating at the second harmonics. The fluorescence emitted upon decay to the final state was measured using a time-correlated photon counting technique, as a function of argon pressure. From this the radiative lifetime was extracted by extrapolating the plot to collision-free zero pressure. We also report calculated radiative lifetimes of the Na$_2$ $6\,^1\Sigma_g^+$ ro-vibrational levels in the range of $v=0-200$ with $J=1$ and $J=31$ using the LEVEL program for bound-bound and the BCONT program for bound-free transitions. Our calculations reveal the importance of the bound-free transitions on the lifetime calculations and a large difference of about a factor of three between the $J=1$ and $J=31$ for the $v=40$ and $v=100$, due to the wavefunction alternating between having predominantly inner and outer well amplitude.
\end{abstract}

\pacs{Valid PACS appear here}


\maketitle

\section{Introduction}
Alkali diatomic molecules have been widely studied by theoreticians and experimentalists for many years due to their simple and rich internal structures. Investigations of molecular structure and dynamics of short-range potentials and photodissociation, predissociation, autoionization, and energy transfer processes using traditional short-range molecular spectroscopy are important for ultracold physics. In recent years cold and ultracold (bi)alkali diatomic molecules have been at the forefront of quantum chemistry and many-body physics. These molecules have been used to probe new states of quantum matter, to improve precision measurements of fundamental constants, and in the development of quantum information storage and molecule lasers~\cite{Stwalley78,Stwalley99,Jones06,Carr09,Bahns00,McGuyer15,Balakrishnan16,Krems08}. While the radiative properties of ultracold molecules are no longer restricted by collisional de-excitation, preparing dense samples of ultracold diatomic molecules in a specific quantum state still remains challenging. Diatomic alkali molecules exhibit an exotic behaviour of the $6\,^1\Sigma_g^+$ symmetry with double well (inner and outer) structures due to the sodium positive and negative ion pair potential energy interaction. The existence of an outer-well ($8a_o<R<40a_o$) of this potential curve may play an important role for the production of cold ground state molecules. This work deals with the experimental and theoretical studies of the $6\,^1\Sigma_g^+$(3s+5s) state in sodium diatomic molecules. \\

The potential curves of electronic states up to the Na$_2$ (3s+5s) dissociation limit for the singlet and triplet electronic states and radiative lifetimes of a band of vibrational levels for various electronic states have been studied using various methods~\cite{Demtroder76,Ducas76,Baumgartner84,Radzewick83,Magnier93,tsai94-2,Sanli15,Anunciado16}. Recent experimental and theoretical studies of the radiative lifetimes of the $2\,^1\Sigma_u^+$ state of sodium dimers have been reported using a molecular beam apparatus and pulsed dye lasers with a pump-delayed probe technique by Anunciado~\textit{et. al.}~\cite{Anunciado16}. Authors demonstrated for the first time the importance of the inclusion of the bound-free transitions into the lifetime calculations. Despite considerable interest in the electronic states of sodium diatomic molecules, the radiative lifetimes of the $6\,^1\Sigma_g^+$ inner-well state have not been measured by experimentalists. The lifetimes of the outer-well of this state have been reported in multi-step laser excitation method~\cite{Laue03}. In this paper we report for the first time lifetime measurements of the inner-well of the $6\,^1\Sigma_g^+ (7,31)$ state using time-correlated photon counting technique and theoretical calculations in various ro-vibrational levels of the $6\,^1\Sigma_g^+$ state using bound-bound and bound-free transitions. We also demonstrate the effect of inclusion of the bound-free transitions in the calculated radiative lifetime of the $6\,^1\Sigma_g^+$ (3s+5s) ro-vibrational levels in the range of $v=0-200$ with $J=1$ and $J=31$. We report the transition dipole moment functions coupling the $6\,^1\Sigma_g^+$ state to seven singlet, ungerade states and also the branching ratios for radiative transitions from the $6\,^1\Sigma_g^+$ ro-vibrational levels into those seven states. We discuss the results of both experimental and theoretical lifetimes of the $6\,^1\Sigma_g^+$  (3s+5s) state.\\

The remainder of this paper is organized as follows: section~\ref{sec:Experiment} contains a description of the experimental apparatus including the excitation scheme and potential energy curves of Na$_2$, section~\ref{sec:Measurements} presents new measurements including the techniques used, section~\ref{sec:Calculations} describes the theoretical calculations, followed by results and discussions in section~\ref{sec:Results}, and conclusion in section~\ref{sec:Conclusion}.

\section{Experiment}\label{sec:Experiment}

\begin{figure*}[th]
\includegraphics[scale=0.85]{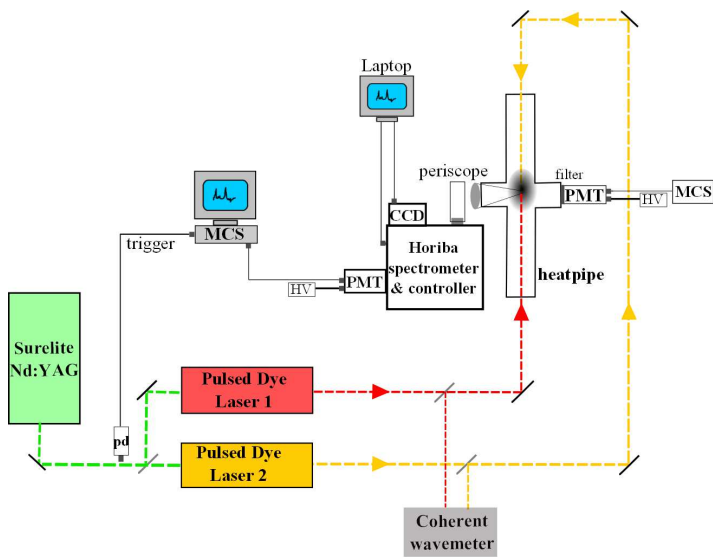}
\caption{(Color online) A schematic view of the experimental setup. The tunable dye lasers are pumped by the pulse Nd:YAG laser. The slit widths and swing-away mirror of the spectrometer are computer-controlled. The CCD is a charge-couple device, cooled to -40~$^\circ$C, with 14x10$^{-6}$ m pixel size. MCS is a multi-channel scaler for time-correlated photon counting, PMT refers to a photomultiplier tube, pd a photodiode, and a periscope, consisting of two mirrors, is used to invert the horizontal image to vertical in order to optimize the fluorescence collection at the entrance slit of the Horiba spectrometer. }\label{fig:Fig1}
\end{figure*}

The time-resolved double-resonance experimental setup is shown in Fig.~\ref{fig:Fig1}. The sodium metal is contained in a 4-arm crossed heat-pipe oven~\cite{Vidal69,Vidal71,Vidal72,Vidal96}. The desired amount of buffer gas was introduced into the heat-pipe when the oven was cold and the the background pressure was about 15 mTorr. The gas valve was closed to keep a fixed amount of argon gas inside the oven. The oven, which has a length of 1 m, is then heated by eight non-magnetic ceramic heaters to produce sodium vapor while cold water is run through copper coils wrapped near the windows of the arms. The heat-pipe oven and the ceramic heaters were wrapped with a ceramic fiber blanket to maintain a homogenous temperature inside the oven. A stainless steel mesh lining the arms of the oven walls acts as a wick to return the liquid metal back to the central region of the oven  in order to prevent condensation onto the windows. During the experiment argon pressure was continuously monitored by an absolute capacitance manometer (MKS 627E), temperature-controlled to 45~$^\circ$C. To create sodium molecules the heat-pipe is heated to 290~$^\circ$C and the temperature of the central region of the oven was monitored with a thermocouple probe. At the operating temperature of 290~$^\circ$C, about 32.6\% of the molecules are in the vibrational ground state $v''=0$, 21.8\% in $v''=1$, 14.6\% in $v''=2$, 9.85\% in $v''=3$,..., 1.45\% in $v''=8$ and less than 1 in levels $v''=9$ and higher.  The Nesmeyanov vapor pressure formula is used to estimate the sodium molecular and atomic vapor densities at the operating temperature of the heat-pipe oven~\cite{Nesmeyanov63}, yielding a number density at 290~$^\circ$C for the sodium dimers of $1.45$x$10^{12}$ cm$^{-3}$ and for the sodium atoms of $7.35$x$10^{14}$ cm$^{-3}$. \\

\begin{figure}[th]
\includegraphics[scale=0.4]{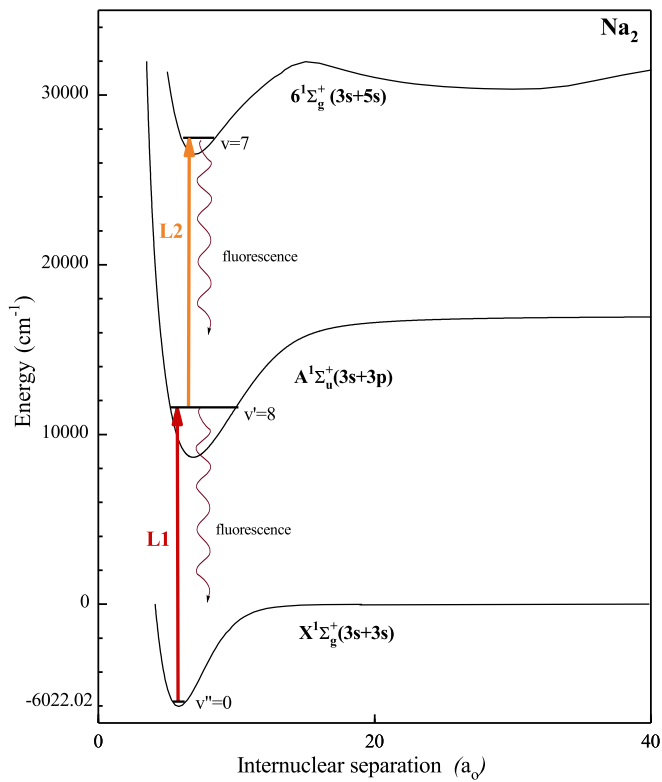}
\caption{(Color online) Experimental excitation to the $6\,^1\Sigma_g^+ (3s+5s)$ state and detection scheme on a partial energy level diagram of Na$_2$.}\label{fig:Fig2}
\end{figure}

The partial energy level diagram of Na$_2$ and experimental excitation scheme are shown in Fig.~\ref{fig:Fig2}. The excitation lasers are tunable home-built pulse dye lasers whose cavity designs are of the grazing incidence Littman-Metcalf~\cite{Littman78} type. Both dye lasers operate in a single transverse mode and are pumped by the second harmonics (532 nm) of a Nd:YAG laser with a pulse repetition rate of 20 Hz. They produce about  0.5 mW average power with bandwidths less than 6 GHz ($\leq$ 0.2 cm$^{-1}$). Dye laser 1, L1,  drives the $X\,^1\Sigma_g^+ (0,29)$ $\rightarrow$ $A\,^1\Sigma_u^+ (8,30)$ transition at 15518.5 cm$^{-1}$ (644 nm) while the second dye laser, L2, drives the $A\,^1\Sigma_u^+ (8,30)$ $\rightarrow$ $6\,^1\Sigma_g^+ (7,31)$ transition at 17798.9 cm$^{-1}$ (561.8 nm). Laser wavelengths are measured using a Coherent wavemeter with a precision of 0.01 cm$^{-1}$. The lasers, both collimated to a 1-cm beam diameter, collinearly counterpropagate in the interaction region of the heat-pipe oven. Pulse duration of the lasers, about 6 ns, was measured using an ultrafast phototube with 270 ps risetime and 100 ps fall time. The L2 pulse arrives at the interaction region of the oven 4 ns after the arrival time of the L1 pulse. Due to the 2 ns temporal overlap time compared to the radiative lifetime of the excited level~\cite{Baumgartner84}, about 23\% of the excited molecules will have radiatively decayed during the pulse. This overlap time sufficiently populates the $6\,^1\Sigma_g^+ (7,31)$ state. Molecular fluorescence emission from the $6\,^1\Sigma_g^+ (7,31)$ state is collected at the right angles to the propagation directions of the lasers using a periscope and an imaging spectrometer-detector system. The imaging spectrometer, Horiba iHR320, acts as a narrowband filter and has three built in diffraction gratings (600, 1200 and 2400 groves/mm), 1200 g/mm is predominantly used throughout the experiment. The spectrometer has two exit ports (each port can be selected with a motorized swing away mirror): the front port has a back illuminated thermoelectrically-cooled (-50~$^\circ$C) charge-couple device (CCD) chip with 14x10$^{-6}$ m squared pixel size while the side port has a slit and a photomultiplier tube (PMT). The widths of the slits and swing away mirror are remotely controlled. The spectrometer is calibrated with a mercury light source. The FWHM resolution of the spectrometer, with the 1200 g/mm holographic grating (blazed at 500 nm) and the side slit at 90x10$^{-6}$ m, is 19.33 cm$^{-1}$ at 17798.9 cm$^{-1}$ and 14.19 cm$^{-1}$ at 15518.5 cm$^{-1}$. For lifetime measurements, the data was collected from the side slit (width at 90x10$^{-6}$ m) using a PMT-multichannel scaler (MCS). A second photomultiplier tube (Hamamatsu R928), with interference filter, was mounted on the second window of the heat-pipe and used to monitor the Na atomic fluorescence from the 4p state to the ground state. \\

\section{Measurements}\label{sec:Measurements}

\begin{figure}[th]
\includegraphics[scale=0.35]{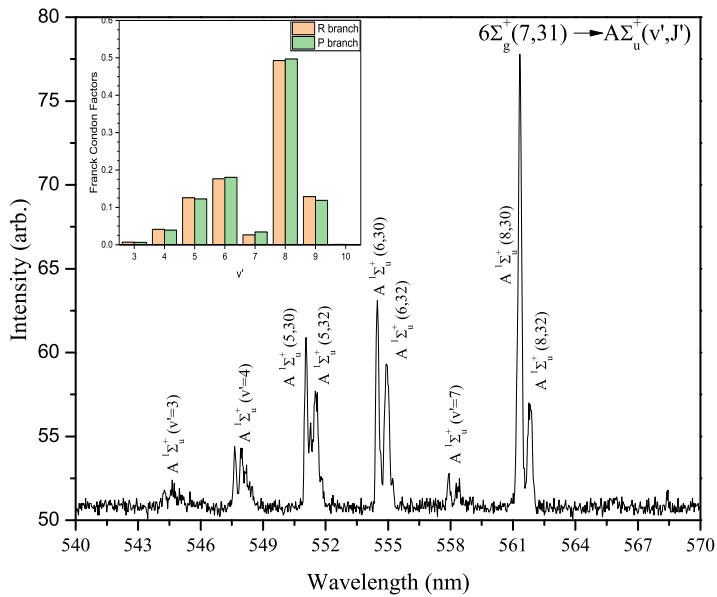}
\caption{(Color online) Double-resonance spectrum from the $6\,^1\Sigma_g^+$ ($7,31$) state. The spectrum, only obtained when both lasers are on resonance, was taken using a 1200 g/mm spectrometer-CCD with FWHM resolution of 3 cm$^{-1}$. The spectral peaks mainly consists of $R(30), P(32)$ doublets for the $A\,^1\Sigma_u^+ (v',J') \leftarrow 6\,^1\Sigma_g^+ (7,31)$ fluorescence emission and they are identified by comparing to the output of the LEVEL 8.0 program. The inset shows the Franck-Condon factors,  extracted from the program, as a function of vibrational quantum number.}\label{fig:Fig3}
\end{figure}

With the setup shown in Fig.~\ref{fig:Fig1}, a resolved molecular fluorescence spectrum was acquired from the $6\,^1\Sigma_g^+ (7,31)$ state as shown in Fig.~\ref{fig:Fig3}. The spectrum exists only when the double-resonance excitation to the  $6\,^1\Sigma_g^+ (7,31)$ state is accomplished. This detection scheme is almost background-free because the frequencies of this spectrum are far away from the molecular fluorescence from the $A\,^1\Sigma_u^+$ state, excited by Laser 1. In addition, it was not possible to observe any molecular fluorescence from the $6\,^1\Sigma_g^+ (7,31)$ state using Laser 1 or Laser 2 alone, and thus no two-photon absorption was observed from the $6\,^1\Sigma_g^+ (7,31)$ state. As a result, the spectrum suggests the two-step double-resonance for the $X\,^1\Sigma_g^+ (0,29)$ $\rightarrow$ $A\,^1\Sigma_u^+ (8,30)$ $\rightarrow$ $6\,^1\Sigma_g^+ (7,31)$ transition. The spectral peaks were identified by comparing them with the output of the LEVEL 8.0 Fortran program. This program uses experimental potential energy curves and calculates quantities such as Franck-Condon factors, Einstein coefficients, transition dipole moment matrix elements and energies of the ro-vibrational levels. High Franck-Condon factors suggest $P-$ and $R-$branches from the $6\,^1\Sigma_g^+ (v=7,J=31)$ to the $A\,^1\Sigma_u^+ (v=3-8, J=30,32)$ state should be readily observed and identified, and they are (see Fig.~\ref{fig:Fig3}). We report the lifetime measurements from the $v=6$ doublets. The separation between the (6,30) and (6,32) doublet is 13.40 cm$^{-1}$ according to the output of the LEVEL fortran program. The photon counting technique is used to perform low-light-level measurement with high sensitivity and accuracy. \\

\begin{figure}[ht]
 {{\includegraphics[scale=0.30]{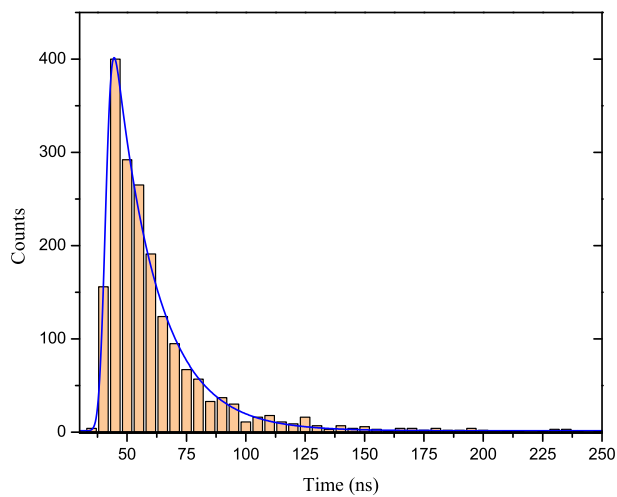} }}%
 \caption{(Color online) A typical time-correlated photon counting scan (40 minutes to complete one decay curve) illustrating a histogram consisting of a range of `time bins' from the MCS. The blue line is a convolution fit of a Gaussian with an exponential curve function to the data with a correlation coefficient of 0.99. The effective lifetime of the molecular excited state was extracted from this convolution fit. }\label{fig:Fig4}
\end{figure}

\begin{figure}[ht]
\includegraphics[scale=0.36]{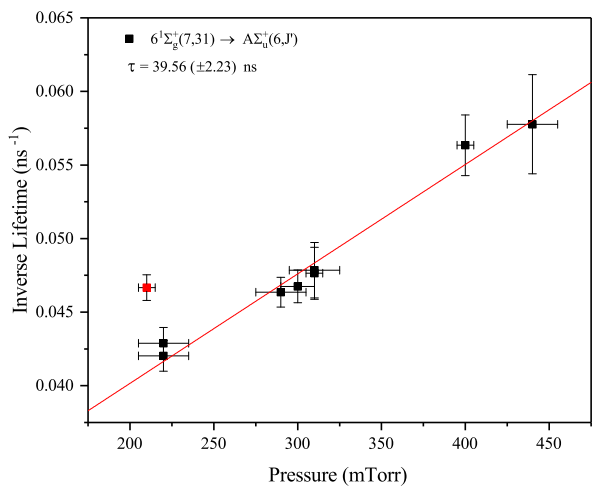}
\caption{(Color online) Plot showing the variation in the inverse lifetime as a function of argon pressure at 290~$^\circ$C. Each data point is an average of 10 to 20 independent lifetime measurements, with the error bar equal one standard deviation. Red filled square data point is an outlier and was excluded in the fitting procedure. Red line shows a weighted linear fit to the data with a correlation coefficient of R$^2$=0.98 and residual sum square of 7.86x10$^{-9}$. }\label{fig:Fig5}
\end{figure}

To do time-correlated photon counting we used a multi-channel scaler (MCS) which has a fast amplifier, a built-in discriminator and 5-ns bin width (the temporal resolution). This technique is used to record low level light signals for the radiative lifetime measurements. The time accuracy is not limited by the width of the PMT detector pulse and thus there is no loss due to gating as in Boxcar devices or gated CCDs. The MCS works by measuring the time between a trigger pulse and an associated PMT pulse which results from a single photon. We measure about one photon per five or more triggers. When the MCS receives a trigger pulse a record of up to about 32,000 time bins, in our case about 1,000 time bins of 5-ns, starts. During a record, when a signal pulse is received, one count is added to the time bin corresponding to when the pulse is received in relation to the trigger. The MCS was triggered by laser pulses from the YAG laser using a photodiode and a scan of 20,000 records was accumulated and displayed on the screen on real-time. The resulting histogram yielded the fluorescence decay as a function of time. If the number of molecules leaving the $6\,^1\Sigma_g^+ (v,J)$ state during the time interval $dt$ is $N=-k_rNdt-k_{nr}Ndt$, the solution gives the emitted radiation with intensity proportional to $N(t)=N_oe^{(k_r+k_{nr})t}$, where $N_o$ is the initial population, $k_r$ is the radiative decay, and $k_{nr}$ is the non-radiative decay. Then, the lifetime is defined as $N(t)e^{-t/\tau_e}$. The typical result, a histogram with an exponential drop of counts as a function of time from the MCS, is shown in Fig.~\ref{fig:Fig4}. The recorded data  were transferred to a computer and the effective lifetime, $\tau_e$, was obtained from the convolution of a Gaussian with an exponential fit to the data using Origin 2017. During the experiment, each scan was accumulated for 40 minutes to produce a full decay curve and more than ten independent measurements were done at one pressure point on various days. The average of the effective lifetime was calculated for each pressure setting. In order to determine the collision-free lifetime, this procedure was repeated at argon pressures ranging from 200 mTorr to 460 mTorr (at the operating temperature) and thus the effective lifetime as a function of pressure was obtained and plotted. A linear dependence of such a plot, known as Stern-Volmer, is expected if the probability of the multiple collisions is negligible. The Stern-Volmer relationship can be written as
\begin{equation}
\frac{1}{\tau_{e}} = \frac{1}{\tau_r} + \frac{1}{\tau_{nr}},
\end{equation} 
where $\tau_{e}$ is the effective lifetime, $\tau_r$ is the radiative lifetime, and $\tau_{nr}^{-1}=N_p\bar{v}\sigma$. Here,  $N_p$ is the number density of the buffer gas, $\bar{v}$ is the average velocities of the colliding Na$_2^*$-Ar atoms over the
Maxwell-Boltzmann distribution of relative velocities at the heat-pipe oven temperature, and $\sigma$ is the collisional cross section between the excited dimer and argon atoms.  The effect of argon gas collisions is eliminated with the Stern-Volmer extrapolation. The radiative lifetime was extracted from the zero pressure extrapolation of the Stern-Volmer plot. 
Figure~\ref{fig:Fig5} shows the inverse lifetimes (ns) as a function of argon pressure at 290~$^\circ$C in the range of 200 mTorr - 450 mTorr. The radiative lifetime at the collision-free limit (intercept of the Stern-Volmer plot) was extracted from a linear fit to the data. The radiative lifetime of the $6\,^1\Sigma_g^+ (7,31)$ state is found to be 39.56 ($\pm$ 2.23) ns with a correlation coefficient of 0.99 and residual sum square of 7.86x10$^{-9}$. \\

The effect of high pressure on the lifetime measurements was observed for pressures from about 500 mTorr to 1700 mTorr. A slight curving trend at high pressures may indicate the possibility of buffer gas quenching effect~\cite{Romalis09} which can be studied in the future. For the purpose of determining the radiative lifetime at the collision-free limit a linear fit to the data was done at low pressures. Additionally, we repeated the time-correlated photon counting experiments at different operating temperatures ranging from 290~$^\circ$C to 340~$^\circ$C and at different laser powers. The data showed no laser power dependency on lifetime measurements. However, as expected, lifetime decreased as we increased the temperature to 340~$^\circ$C. To ensure the reproducibility of the data we have repeated the experiments on different days. We have validated our measurements and setup by repeating the measurements for the known lifetimes of the Na$_2$ $A\,^1\Sigma_u^+ (7,31)$ state under the same experimental conditions. The extracted radiative lifetime from the Stern-Volmer plot is found to be 14.02 (93) ns, comparable with the literature values of 13.11 (47) ns~\cite{Ducas76} and 12.51 (62) ns~\cite{Baumgartner84}. \\

In addition, we observed atomic fluorescence from the  Na $4p\,^{2}P_{J}$ state to the $3s\,^{2}S_{1/2}$ ground state using a 330 nm interference filter.  This is due to the molecular dissociation into the Na$_2$ (3s+4p) asymptote. The signal exists only when molecules were formed in the heat-pipe. To ensure that the atomic fluorescence is not a result of an accidental dissociation into the Na$_2$ (3s+5s) asymptote and then branching into the $4p$ state we monitored fluorescence for the  $5s\,^{2}P_{J} \rightarrow 4p\,^{2}P_{J}$ transition at 615 nm and found no evidence.

\section{Theoretical Calculations}\label{sec:Calculations}

\begin{figure}[th]
\includegraphics[scale=0.4]{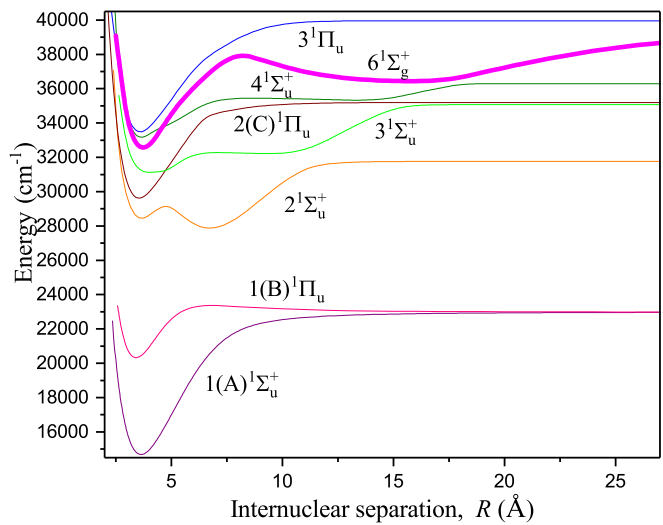}
\caption{(Color online) Na$_2$ $6\,^1\Sigma_g^+$ potential energy function and the singlet, ungerade electronic states used in calculations into which excited molecules in the $6\,^1\Sigma_g^+$ state can decay.}\label{fig:Fig1S}
\end{figure}

\begin{figure}[th]
\begin{center}
\includegraphics[scale=0.4]{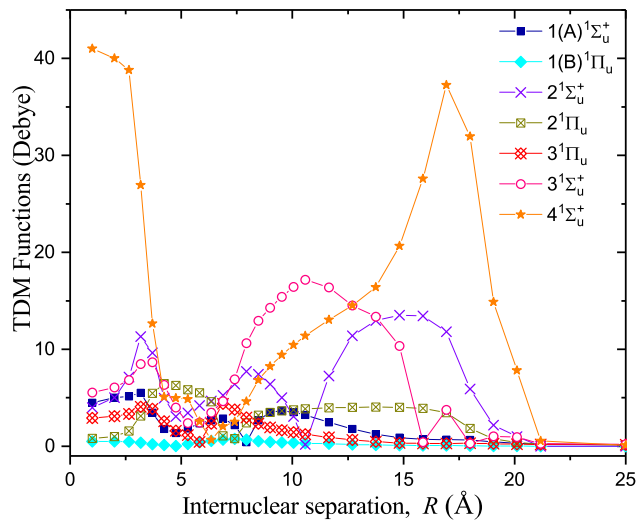}
\caption{(Color online) Transition dipole moment (TDM) functions that couple the $6\,^1\Sigma_g^+$ state to each of the other potentials used in the calculations.}\label{fig:Fig2S}
\end{center}
\par
\end{figure}

All lifetime calculations presented here were performed using the LEVEL 8.2 Fortran program, developed by Le Roy~\cite{LeRoy8.2}, for the bound-bound transitions and the BCONT Fortran program for the bound-free transitions. The LEVEL code solves the radial Schr{\"o}dinger equation of bound and quasibound levels with appropriate input for the transition dipole moment function~\cite{Magnier17} and relevant potential energy curves. The dipole selection rules allow spontaneous emission from the $6\,^1\Sigma_g^+ (v,J)$ (Tsai~\emph{et al.}~\cite{tsai94-2} and Laue~\emph{at al.}~\cite{Laue03}) to seven electronic potentials: $1(A)\,^1\Sigma_u^+$ from Tiemann~\emph{et al.}~\cite{Tiemann96}, $1(B)\,^1\Pi_u^+$ from Comacho~\emph{et al.}~\cite{Camacho05} and Tiemann~\emph{et al.}~\cite{tiemann87}, $2\,^1\Sigma_u^+$ from Pashov \emph{et al.}~\cite{Pashov00-1}, $2\,^1\Pi_u^+$ from Grochola \emph{et al.}~\cite{Grochola05}, $3\,^1\Sigma_u^+$~\cite{Sanli18}, $3\,^1\Pi_u^+$ from Grochola \emph{et al.}~\cite{Grochola06}, and $4\,^1\Sigma_u^+$ from Grochola \emph{et al.}~\cite{Grochola04}. The potentials and transition dipole moment functions for the seven dipole allowed transitions from the $6\,^1\Sigma_g^+ (v,J)$ state are shown in Fig.~\ref{fig:Fig1S} and Fig.~\ref{fig:Fig2S}.\\

The LEVEL 8.2 program computes the Einstein $A$ coefficient for spontaneous emission, $A_{if}$, from an initial level $|i\rangle$ = $|\alpha,v,J\rangle$ to a final level $|f\rangle$ = $|\alpha{'},v',J'\rangle$. This coefficient can be expressed as~\cite{Herzberg50,Bernath16}
\begin{equation}\label{eq:A}
A_{if}=\frac{16\pi^3}{3}\frac{k_{if}^3}{\epsilon_oh}\frac{S_{JJ'}}{(2J+1)}|\langle\psi_{v',J'}^{\alpha{'}}|M(r)|\psi_{v,J}^\alpha\rangle|^2,
\end{equation}
where $\alpha$ denotes the electronic state, $v$ and $J$ are vibrational and rotational quantum numbers respectively, $A_{if}$ has units $s^{-1}$, $M(r)$ is the transition dipole moment function, $k_{if}$ the emission wavenumber, $S_{JJ'}$ the H{\"o}nl-London rotational intensity factor, $\psi_{v,J}^\alpha$ is the normalized radial nuclear wave functions belonging to initial and final state of the transition.\\

\begin{figure}[h]
\begin{center}
\includegraphics[scale=0.35]{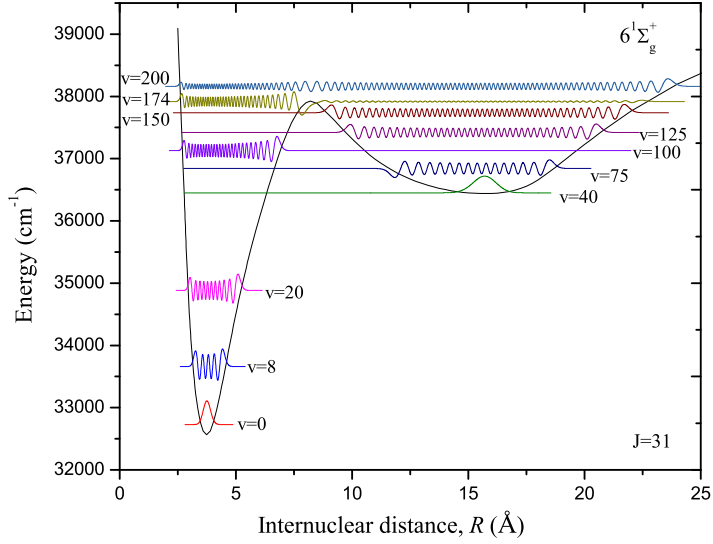}
\caption{(Color online) $6\,^1\Sigma_g^+$ potential with the wavefunctions of the $v=0, 8, 20, 40, 75, 100, 125, 155, 174$, and $200$, all with $J=31$, plotted at the energy of their respective ro-vibrational level.  Note that the wavefunctions  existing in the region where two separated wells exist ($v=40, 75, 100, 125, 155$, and $174$) do oscillate in both wells, however all but $v=174$ must be scaled up several orders of magnitude for those oscillations to become visible. }\label{fig:Fig3S}
\end{center}
\end{figure}

In cases where the sum of the Franck-Condon factors is less than one, we incorporated bound-free transitions using BCONT code  which computes the intensity as a function of wavelength. All bound-free calculations were completed using a modified version of the BCONT 2.2~\cite{LeRoy2.2} program. The modifications have been described previously in Ref.~\cite{Anunciado16} and even greater detail in Ref.~\cite{Brett10} and therefore will not be repeated here. The bound-free calculations are more labor intensive than the bound-bound transitions, and therefore the ro-vibrational levels $6\,^1\Sigma_g^+ (v= 0, 7, 8, 20, 40, 75, 100, 125, 155, 174, 200, J = 1, 31)$ were selected because they include energy levels with wavefunctions that reside in the inner well below the double well region, in the double minimum region, and in the region just above where the inner and outer well merge. Figure~\ref{fig:Fig3S} shows some of the calculated wave functions used by the LEVEL and BCONT programs.\\

\begin{figure}[th]
\includegraphics[scale=0.3]{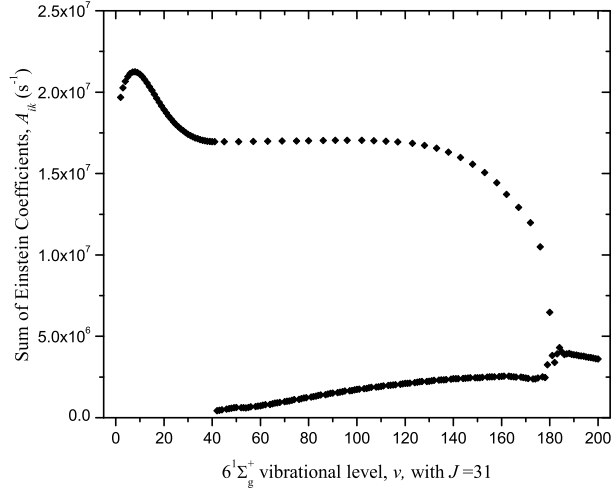}
\caption{Each point on this plot represents the sum of all Einstein coefficients originating from a single ro-vibrational level, $6\,^1\Sigma_g^+ (v,J)$, to all bound levels of final state $\alpha^\prime$. In this plot, the final state is $\alpha^\prime = 1(A)\,^1\Sigma_g^+$ and the vibrational levels $v=0-200$ with $J=31$ are shown.  The upper branch is produced by  $6\,^1\Sigma_g^+ (v,J)$ levels with predominantly inner well wavefunctions, while the lower branch corresponds to those with predominantly outer well wavefunctions.  Around approximately $v=180$, the wavefunctions accrue substantial amplitude in both the inner and outer region as the two wells merge.}\label{fig:Fig4S}
\end{figure}

\begin{figure*}[ht]
 \subfloat[]{{\includegraphics[scale=0.35]{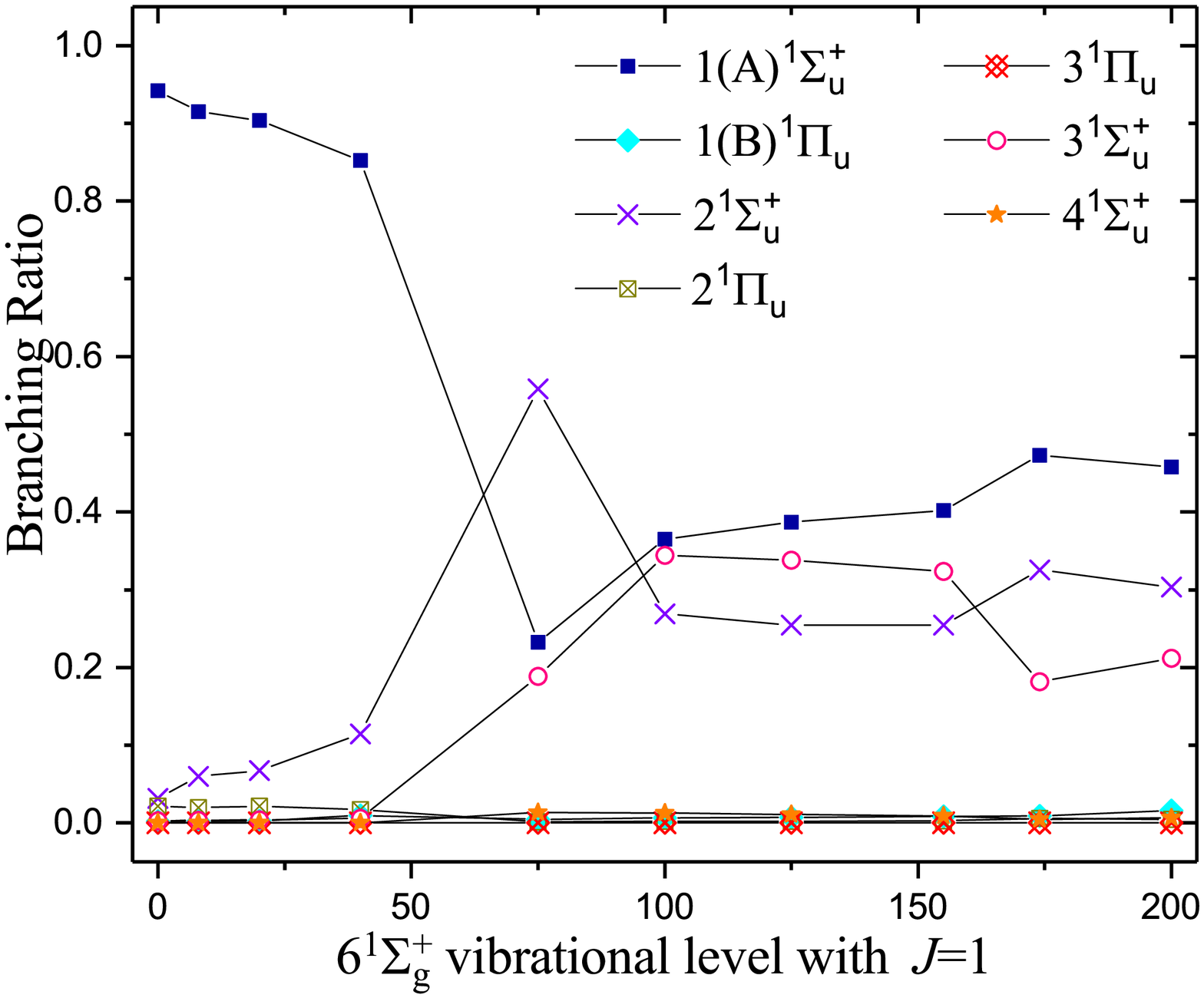} }}%
 \subfloat[]{{\includegraphics[scale=0.35]{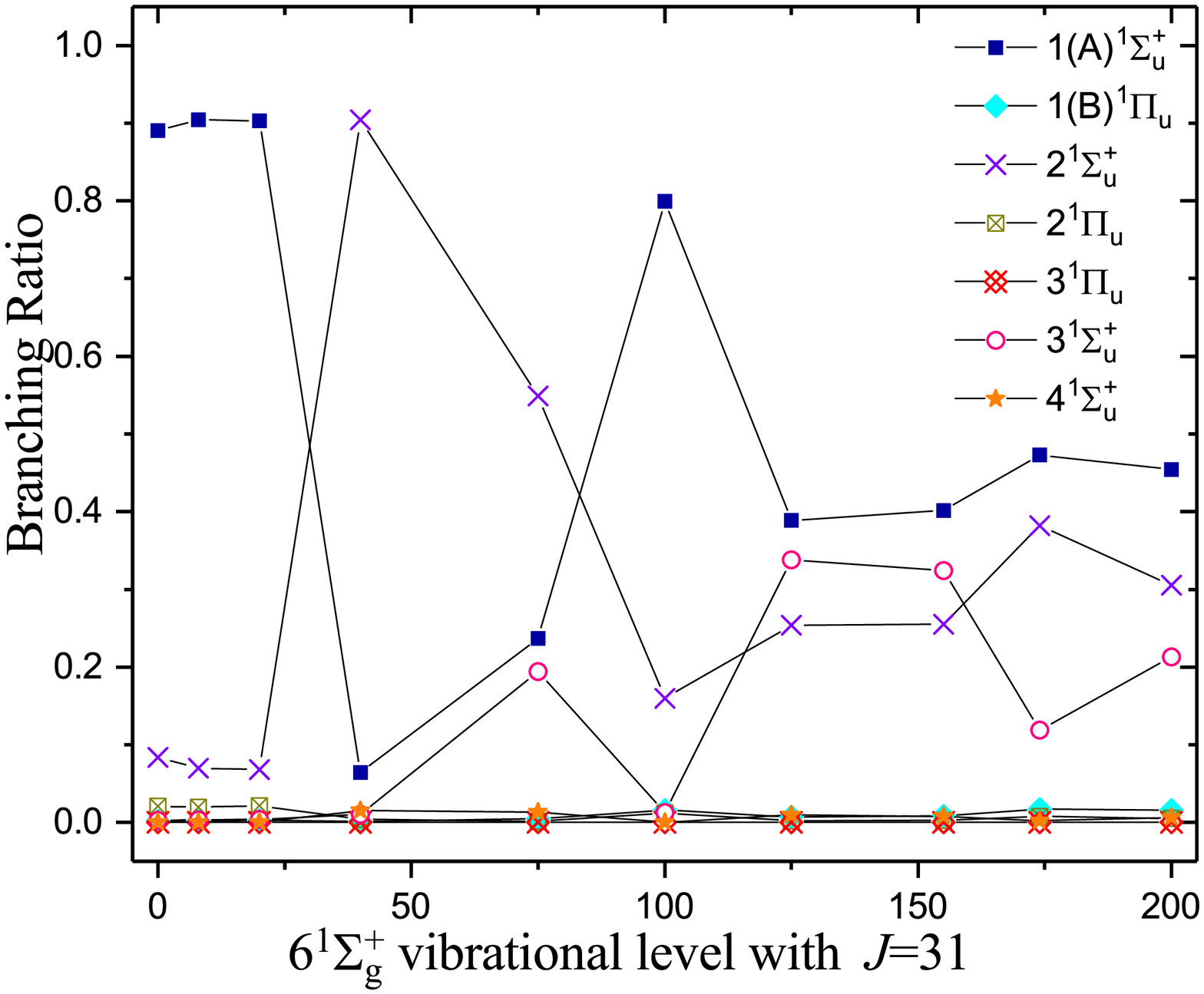} }}%
 \caption{(Color online) Branching ratios for $6\,^1\Sigma_g^+$ ro-vibrational levels with $J$=1 and $J$=31.}\label{fig:Fig6S}
\end{figure*}

The revised BCONT code calculates the bound-free intensity $dI$ per molecule emitted into a small energy band $dE$ at photon energy $E=hck$ given by $dI = E \cdot \Gamma_{if}^{\alpha{''}}dE$, where the photon emission rate is given by
\begin{equation}\label{eq:Gamma}
\Gamma_{if}^{\alpha{'}}dE=\frac{16\pi^3}{3}\frac{k_{if}^3}{\epsilon_oh}\frac{S_{JJ'}}{(2J+1)}|<\psi_{v',J'}^{\alpha{'}}|M(r)|\psi_{v,J}^\alpha>|^2dE
\end{equation}
The main difference between this photon emission rate and the bound-bound version in Eq.~(\ref{eq:A}) is the replacement of the final state bound wave function $\psi_{v,J}^\alpha$ with the continuum wave function, $\psi_{EJ}^\alpha$. The lifetime of a single ro-vibrational level of the  $6\,^1\Sigma_g^+$ is determined using the relation
\begin{equation}\label{eq:tau}
\frac{1}{\tau_i}=\sum_{\alpha{'}}\bigg(\sum_{f}A_{if}+\int_{0}^{E(\alpha{'})}\Gamma_{if}^{\alpha{'}}dE\bigg),
\end{equation}
where the $A_{if}$ coefficients have been obtained from the LEVEL calculations, the $\Gamma_{if}^{\alpha{'}}dE$ from BCONT, and the overall sum includes all dipole allowed transitions. The bound-bound sum extends over all levels in the given electronic state $\alpha^\prime$, and the integral extends from zero frequency to the appropriate threshold $E(\alpha{'})=E(v,J)-E(\alpha^\prime$ asymptotic energy$)$. The sum of Einstein coefficients as a function of $6\,^1\Sigma_g^+$ vibrational level for transitions to the $A\,^1\Sigma_u^+$ is plotted in Fig.~\ref{fig:Fig4S}. \\
If the $A_{if}$ and $\Gamma_{if}^{\alpha^\prime}dE$ for a single final electronic state, $\alpha^\prime$, are summed and divided by the full sum over all $\alpha^\prime$ in Eq.~\ref{eq:tau}, one obtains the branching ratio, $r_b^\alpha$, into final state $\alpha^\prime$, i.e.,
\begin{equation}
r_b^a=\frac{(\sum_{f}A_{if}+\int_{0}^{E(\alpha{'})}\Gamma_{if}^{\alpha{'}}dE)}{\sum_{{\alpha{'}}}(\sum_{f}A_{if}+\int_{0}^{E(\alpha{'})}\Gamma_{if}^{\alpha{'}}dE)}.
\end{equation}
Branching ratios for $J$=1 and $J$=31, for the calculations that include both bound-bound and bound-free transitions, are plotted in Fig.~\ref{fig:Fig6S}. Note the rather small branching ratios for the $1(B)\,^1\Pi_u$, $2\,^1\Pi_u$, $3\,^1\Pi_u$, and $4\,^1\Sigma_u^+$ states indicate that they have minimal impact on the resulting lifetimes, since the excited molecule has a lower probability of decaying through those states. Conversely, the $1(A)\,^1\Sigma_u^+$, $2\,^1\Sigma_u^+$, and $3\,^1\Sigma_u^+$ states offer the most probable decay paths for the excited molecule.\\

\begin{figure*}[ht]
\centering
 \subfloat[]{
 {\includegraphics[scale=0.3]{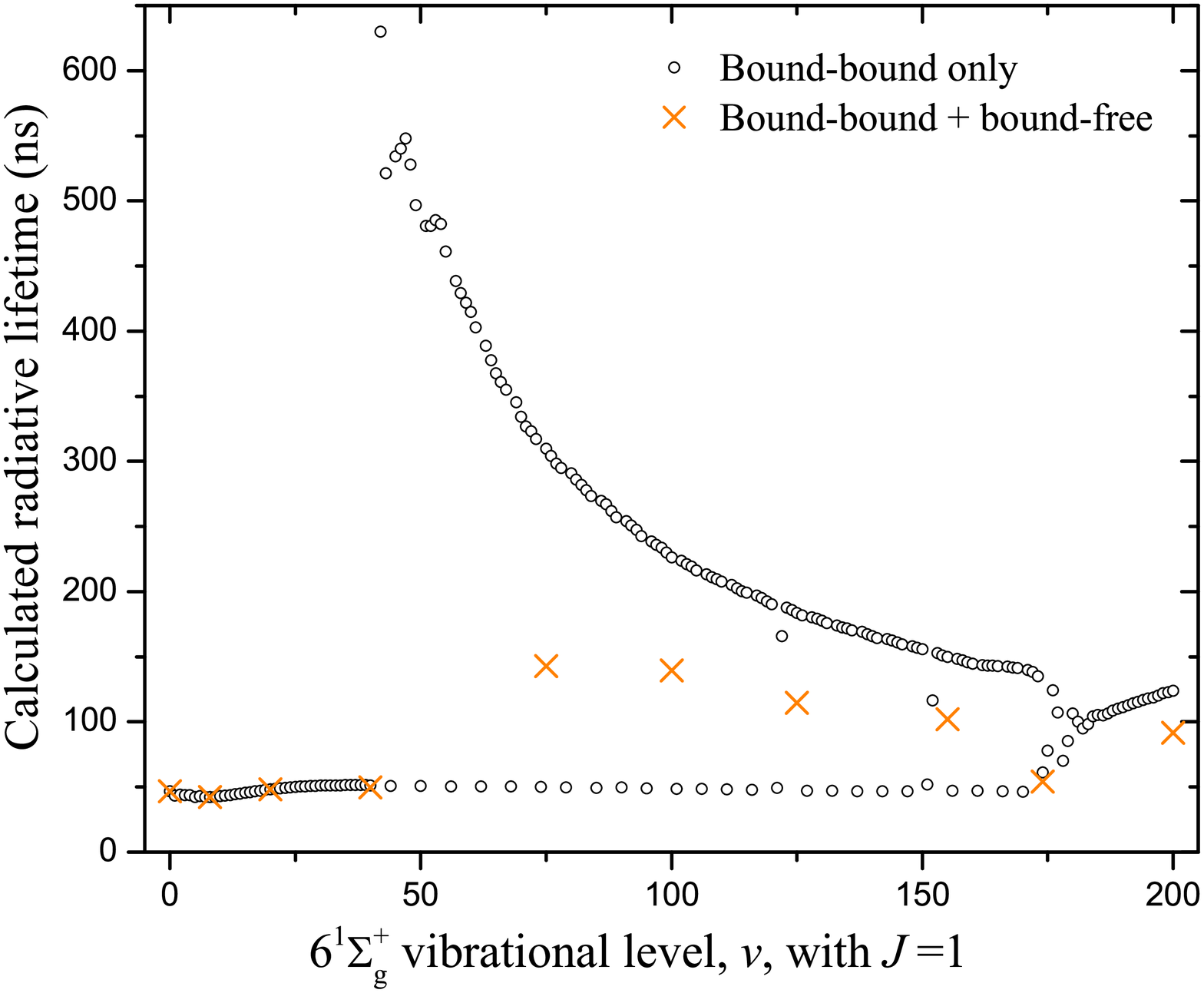} }
 }%
 \subfloat[]{{\includegraphics[scale=0.3]{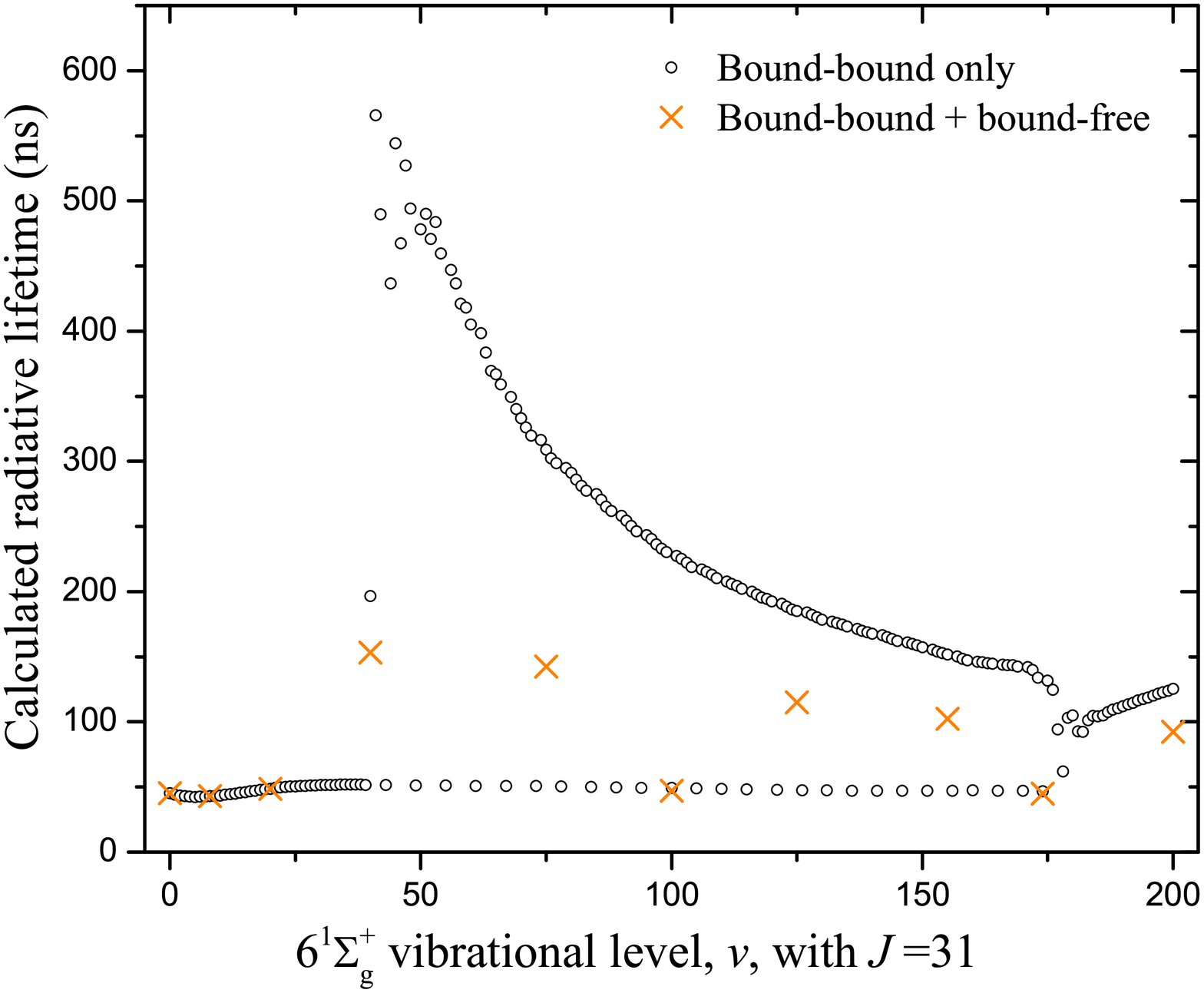} }
 }%
 \caption{(Color online) Calculated lifetimes of $6\,^1\Sigma_g^+(v,J=1,31)$ ro-vibrational levels in the range $v=0-200$. The upper branch is produced by ro-vibrational states with wavefunctions that have predominantly outer well amplitude, while the lower branch wavefunctions are predominantly inner well. The open circles were computed using only bound-bound transitions, $\tau^{-1}=\sum_{\alpha{'}}\sum_{f}A_{if}^{\alpha{'}}$, while the orange X represents lifetimes calculated as in Eq.~(\ref{eq:tau}). The inclusion of bound-free transitions causes a much larger reduction of the calculated lifetimes for the ro-vibrational levels with appreciable outer well wavefunction amplitude since there is generally a less favorable overlap of the $6\,^1\Sigma_g^+$ outer well.}\label{fig:Fig7S}
 \end{figure*}

\begin{figure}[ht]
\centering
\includegraphics[scale=0.4]{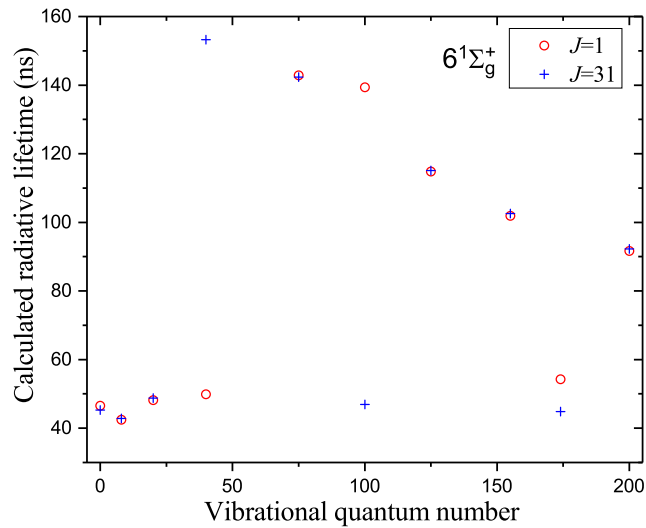}
\caption{(Color online) Comparison of radiative lifetimes of J=1 and J=31 calculated for selected vibrational levels $v=0,8,20,40,75,100,125,155,174,$ and $200$, using Eq.~(\ref{eq:tau}).}\label{fig:Fig9S}
\end{figure}

Figure~\ref{fig:Fig7S} shows our calculated radiative lifetimes of the $6\,^1\Sigma_g^+$ ro-vibrational levels in the range of $v=0-200$ with $J=1$ and $J=31$. This figure clearly demonstrates the importance of the bound-free transitions in the calculations.
A strong dependence of the calculated radiative lifetime on the radial coordinate, $R$, is observed, causing two distinct branches to occur in the portion of the $6\,^1\Sigma_g^+$   potential in which the two wells remain separated. Figure~\ref{fig:Fig9S} shows the $6\,^1\Sigma_g^+$   radiative lifetimes obtained from the full (Eq.~\ref{eq:tau}) calculation for $J=1$ and $J=31$, plotted as a function of their vibrational level.  The large difference of approximately a factor of three between the $J=1$ and $J=31$ lifetimes for $v=40$ and for $v=100$ occurs due to a ``switch'' of the wavefunction, which is predominantly outer well for the long lifetime value, $v=40,J=31$ and $v=100,J=1$, and is predominantly inner well for the shorter lifetime value, $v=40,J=1$ and $v=100,J=31$. The results of these calculated radiative lifetimes are presented in Table~\ref{tab:table1}. \\

\begin{table}[th!]
\caption{\label{tab:table1}Calculated radiative lifetimes (in nanosecond) of selected ro-vibrational levels of the Na$_2$ $6\,^1\Sigma_g^+$ state.}
\begin{ruledtabular}
\begin{tabular}{ccc}
$v$  &  $J=1$  & $J=31$  \\
\hline
0 & 46.4 & 45.3\\
7 & 42.5 & 42.8\\
8 & 42.5 & 42.9\\
20 & 48.2 & 48.8\\
40 & 49.9 & 153.2\\
75 & 142.9 & 142.4\\
100 & 139.4 & 46.9\\
125 & 114.8 & 115.1\\
155 & 101.9 & 102.6\\
174 & 54.3 & 44.9\\
200 & 91.7 & 92.3\\
\end{tabular}
\end{ruledtabular}
\end{table}

\section{Results and Discussions}\label{sec:Results}

A time-resolved double resonance spectroscopy technique has been applied to the sodium diatomic molecules in a heat-pipe and the radiative lifetime of the $6\,^1\Sigma_g^+ (7,31)$ state was measured using a time-correlated photon counting technique. Also, lifetime calculations in various ro-vibrational levels of the $6\,^1\Sigma_g^+$ state using bound-bound and bound-free transitions were performed.\\

The radiative lifetimes are measured and the effect of argon gas collisions is eliminated with the Stern-Volmer extrapolation, explained in section~\ref{sec:Measurements}. The measured radiative lifetime (extracted from the zero-pressure of the extrapolation) of the $6\,^1\Sigma_g^+$ ($7,31$) state is found to be 39.56~($\pm$2.23) ns and our calculations showed this value to be 42.8 ns.  Experimental and calculated radiative lifetimes reported in this work have been compared and the results were in reasonable agreement. To validate our measurements and the techniques, we also measured the radiative lifetime of the $A\,^1\Sigma_u^+ (8,30)$ state by extrapolating the Stern-Volmer plot to zero pressure. The result is in good agreement with the literature values within the experimental uncertainties. \\

The observation of the non-linear behaviour on the lifetimes at high pressures in the range of 500 mTorr - 1800 mTorr may be due to the quenching effect. Further experimental and theoretical studies of the high pressure effect on the lifetime will be investigated in the future work. For the purpose of determining the radiative lifetime at the collision-free limit a linear fit to the data was done at pressures below 500 mTorr. In addition, the observed dissociation signal from the Na 4p state does not originate from any resonance transition to the $6\,^1\Sigma_g^+$ ($7,31$) state. The measured spectrum exists only when both lasers are turned on (see Fig. 3).  Since the $6\,^1\Sigma_g^+$ ($7,31$) state is about 2500 cm$^{-1}$ below the (3s+4p) asymptote collisional energy ($kT\approx$ 600~cm$^{-1}$) is not sufficient to transfer $6\,^1\Sigma_g^+$ ($7,31$) molecules into this asymptote. In addition, dissociation signal appears only when molecules are formed in the heat-pipe ($\sim 280~^\circ$C), indicating that no atomic transitions exist with either laser. As a result, the observed dissociation signal has no effect on the measured lifetime. A possible channel for the dissociating molecule into the (3s+4p) asymptote may be through a one-color two-photon transition. The information about the origin of observed dissociation can be studied in detail in a separate work and will not be detailed here.\\

Table~\ref{tab:table2} is the tabulated outline of the measurements and theoretical calculations. A reasonable agreement between the experimental and theoretical results are achieved when bound-bound and bound-free transitions are considered.
\begin{table}[H]
\caption{\label{tab:table2}Comparison of measured and calculated radiative lifetimes of selected ro-vibrational levels of the Na$_2$ $6\,^1\Sigma_g^+$. Theoretical calculations are compared according to the bound-bound, denoted as bb, and bound-free, denoted as bf, transitions.}
\begin{ruledtabular}
\begin{tabular}{ccccc}
Method  &  Level & $\tau~$(ns) & Reference\\
\hline
Experiment & $6\,^1\Sigma_g^+ (7,31)$ & $39.56~(\pm 2.23)$ & This work\\
Theory~(bb+bf) & $6\,^1\Sigma_g^+ (7,31)$ & $42.8$                  & This work\\
Theory~(bb+bf) & $6\,^1\Sigma_g^+ (7,1)$ & $42.5$                   & This work\\
Theory~(bb)& $6\,^1\Sigma_g^+ (7,1)$   & $126$                  & Ref.~15\\
\end{tabular}
\end{ruledtabular}
\end{table}

\section{Conclusion}\label{sec:Conclusion}

For the first time, experimental lifetime measurement of the $6\,^1\Sigma_g^+ (7,31)$ inner well and theoretical calculations in various ro-vibrational levels of the $6\,^1\Sigma_g^+$ state using bound-bound and bound-free transitions were performed. In addition, the effect of inclusion of the bound-free transitions in the calculated radiative lifetime of the $6\,^1\Sigma_g^+$ (3s+5s) ro-vibrational levels in the range of $v=0-200$ with $J=1$ and $J=31$ were reported. The results reveal the importance of the bound-free transitions and rotational quantum number (e.g. a large difference of about a factor of three between the $J$ = 1 and $J$ = 31 for the $v$ = 40 and $v$ = 100) on the lifetime calculations. Also, the transition dipole moment functions coupling the $6\,^1\Sigma_g^+$ state to seven singlet, ungerade states and also the branching ratios for radiative transitions from the $6\,^1\Sigma_g^+$ ro-vibrational levels into those seven states were demonstrated. The measured and calculated radiative lifetimes are found to be 39.56~($\pm$ 2.23) ns and 42.8 ns, respectively. The results are in reasonable agreement.

\section{Acknowledgements}
Financial support from the National Science Foundation (Grant No. NSF-PHY-1607601) is gratefully acknowledged. Also, we acknowledge Professor Sylvia Magnier of Universite Lille, France for providing transition dipole moment functions used in our calculations. Authors from Miami University thank Professor Mark Havey of Old Dominion University for lending the MCS and Professor Marjatta Lyyra of Temple University for the heat-pipe.



\end{document}